\documentclass[amssymb,amsmath]{article}
\usepackage{graphicx}
\usepackage{dcolumn}
\usepackage{bm}
\usepackage{natbib}

\usepackage{graphicx}

\newcommand{\Rtwo}{{\mathrm R_{\mathrm 2}}}  
\newcommand{\Stwo}{{\mathrm S_{\mathrm 2}}}  
\newcommand{\Atwo}{{\mathrm A_{\mathrm 2}}}  
\newcommand{\Btwo}{{\mathrm B_{\mathrm 2}}}  
\newcommand{\transmA}{\nu_{m_A}}		
\newcommand{\transmB}{\nu_{m_B}}		

\newcommand{\transmR}{\nu_{m_R}}		
\newcommand{\transmS}{\nu_{m_S}}		

\begin{document}


\title{Designing sequential transcription logic: a simple genetic circuit for conditional memory}
\date{}
\maketitle 
\author{Georg Fritz$^{1,2}$},
\author{Nicolas E. Buchler$^3$},
\author{Terence Hwa$^4$}, and
\author{Ulrich Gerland$^{1}$}
 \\ 
 \begin{center}
{\small
$^1$Institute for Theoretical Physics, Universit\"at zu K\"oln, Z\"ulpicher Str. 77, 50937 K\"oln, Germany\\
$^2$Department of Physics and CeNS, LMU M\"unchen, Geschwister-Scholl-Platz 1, 80339 M\"unchen, Germany\\
$^3$Center for Studies in Physics and Biology, The Rockefeller University, New York, NY 10021\\
$^4$Center for Theoretical Biological Physics and Physics Department, University of California at San Diego, La Jolla, CA 92093-0374\\ 
}
\end{center}

\begin{abstract} 
The ability to learn and respond to recurrent events depends on the capacity to remember transient biological signals received in the past. Moreover, it may be desirable to remember or ignore these transient signals conditioned upon other signals that are active at specific points in time or in unique environments. Here, we propose a simple genetic circuit in bacteria that is capable of conditionally memorizing a signal in the form of a transcription factor concentration. The circuit behaves similarly to a ``data latch'' in an electronic circuit, i.e. it reads and stores an input signal only when conditioned to do so by a ``read command''. Our circuit is of the same size as the well-known genetic toggle switch (an unconditional latch) which consists of two mutually repressing genes, but is complemented with a ``regulatory front end'' involving protein heterodimerization as a simple way to implement conditional control. Deterministic and stochastic analysis of the circuit dynamics indicate that an experimental implementation is feasible based on well-characterized genes and proteins. It is not known, to which extent molecular networks are able to conditionally store information in natural contexts for bacteria. However, our results suggest that such sequential logic elements may be readily implemented by cells through the combination of existing protein-protein interactions and simple transcriptional regulation. 
\end{abstract}


\newpage
\section{Introduction}

Biological information is primarily stored in the nucleotide sequence of the genomic DNA. However, cells also use other, epigenetic forms of storage such as DNA methylation patterns, histone modifications, and gene expression states \citep{Jaenisch_Nat_Genet_03,Casadesus_Bioessays_02}. The information contained in these forms of biochemical memory is selectively modified by signals during the life cycle of a cell, yet is stable so that cells can pass on this memory to their daughter cells for many generations. For instance, such memory plays an important role in development, where cell differentiation signals lead to chemical modification of chromatin \citep{Arney_J_Cell_Sci_04}. Signals in development are often transient and the decision to remember or ignore these events can be conditioned upon the presence of hierarchical signals. Such a conditional memory is interesting also for the design of synthetic gene circuits \citep{Basu_Nature_05}. Here, we consider a genetic circuit, which implements conditional memory in its simplest form: it will either store or ignore the signal, depending on the state of a distinct ``read signal''. In this circuit, the storage of memory is implemented via protein concentrations analogous to genetic switches in temperate phages \citep{Trusina_PLoS_Comput_Biol_05}. 

\begin{figure}[h]
\centerline{\includegraphics[width=12cm]{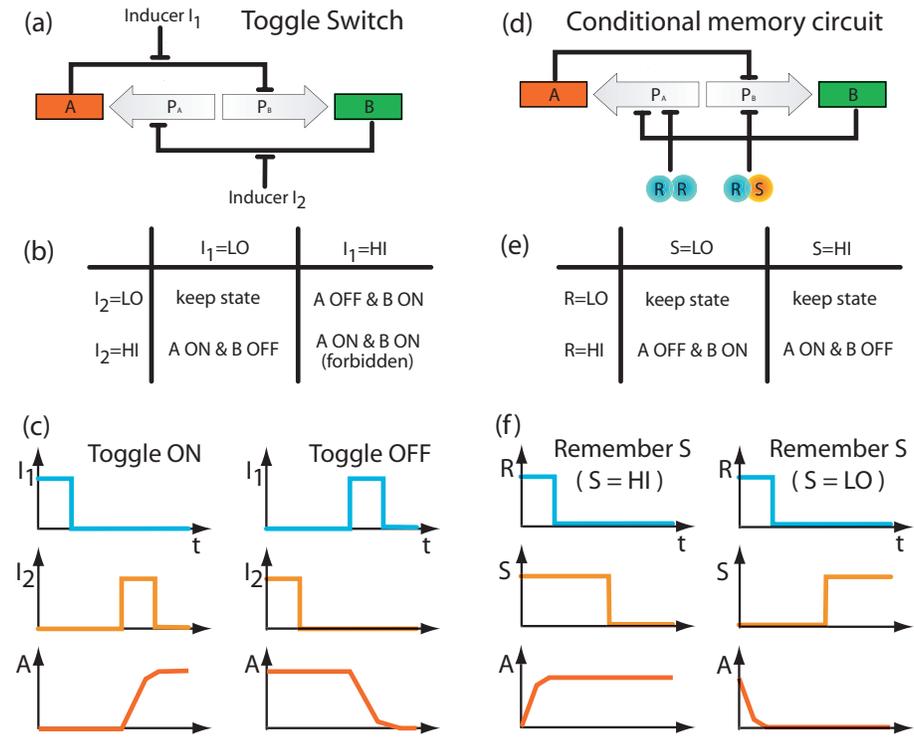}} 
\vspace*{0.3cm}
\caption{\label{FIGscheme} 
Design and Function of the Toggle Switch Compared to the Conditional Memory Circuit.
}
\end{figure}

Our starting point is the synthetic toggle switch of \cite{Gardner_Nature_00}, which exhibits bistability. It consists of two genes, A and B, which mutually repress each other so that they can be in two stable expression states with either A high, B low or A low, B high, see Fig.~\ref{FIGscheme}(a). The toggle switch is coupled to two inducers, which set the switch to the ON-state (A high) or reset it to the OFF-state (A low), see Fig.~\ref{FIGscheme}(b) and (c). The memory of this toggle switch is {\em unconditional}, in that it always stores the state set by the latest inducer exposure. To obtain a conditional memory, we introduce a ``regulatory front end'' to the toggle switch of Fig.~\ref{FIGscheme}(a), through the combinatorial control by two other proteins, R and S, see Fig.~\ref{FIGscheme}(d). The protein R can form homodimers $\Rtwo$ and repress the transcription of gene A. It can also bind to S to form heterodimers RS which repress the transcription of gene B. Qualitatively, we expect that if R is absent, neither $\Rtwo$ nor RS can form: in this state, the existing memory of the circuit is maintained regardless of the state of the level of S. When a significant amount of R is present, it will mostly form homodimers $\Rtwo$ at low concentrations of S, so that gene A is repressed and the switch is forced into the OFF state. Conversely, when S is highly expressed, mostly the heterodimers RS will form and force the switch into the ON state, see Fig.~\ref{FIGscheme}(e) and (f). Hence, at a high level of R, the state of gene A should reflect the state of S. In the language of electrical circuits, a high level of R is the command (or condition) for the system to ``read'' the input signal (S), which is then ``memorized'' when R is subsequently set to a low level. 


Indeed, \cite{Hollis_PNAS_88} characterized a pair of proteins, the 434 repressor and its mutant $434R[\alpha 3(P22R)]$, which behaves like R and S described above. The concept of a regulatory front end has been experimentally proven by \cite{Kobayashi_PNAS_04}, who interfaced several `sensor modules' to the toggle switch. Incidentally, both the basic toggle switch and the conditional memory circuit are examples of what in digital circuit theory is called {\em sequential logic} \citep{Katz_04}. In contrast to {\em combinatorial logic}, the output of which is only a function of its present input signals, sequential logic has an output that depends also on the history of the input signals. In digital circuits, sequential logic is used to construct delay and storage elements, and so-called finite state machines. Indeed, a typical digital circuit consists of both combinatorial and sequential logic. Analogously, one might expect that ``computation'' in natural genetic networks is also implemented by motifs of combinatorial and sequential logic elements. We have previously studied aspects of combinatorial logic in transcription control \citep{Gerland_PNAS_02,Buchler_PNAS_03}. The present study is an attempt to characterize similar aspects of sequential logic in gene regulation, which may be used in natural or synthetic genetic circuits. In particular, we use the conditional memory circuit to illustrate that the combination of protein-protein interactions and {\it cis}-regulatory protein-DNA interactions can supply a particularly simple and compact way to implement complex sequential logic, without the need for additional genetic circuitry. 

The paper is organized as follows. We first verify the basic functionality outlined above within a simple and transparent gene expression model, and characterize the working principle of the circuit from a dynamical systems perspective. Within a more detailed, stochastic model, we then study the effects of biochemical noise on the functionality of the circuit and discuss the experimental feasibility. For the toggle switch of \cite{Gardner_Nature_00}, the sole effect of stochasticity is the spontaneous flipping of the switch \citep{Kepler_BiophysJ_01,Aurell_PRL_02}. We will see that the conditional memory circuit has an additional source for noise-induced errors during the time periods where the read signal is presented. Finally, in the discussion section, we assess the function of the conditional memory circuit both from a biological and a digital circuit perspective. 


\section{Results}

\subsection{Illustration of the circuit function with a reduced model}

The working principle of the conditional memory circuit of Fig.~\ref{FIGscheme}(d) is best understood within a reduced deterministic description, which considers explicitly only the time evolution of the total concentrations of the proteins A and B. Such a description assumes that all biochemical processes which do not change the total concentrations $A_{\rm tot}$ and $B_{\rm tot}$ are so rapid that they remain equilibrated at almost all times. The net change of $A_{\rm tot}$ and $B_{\rm tot}$ due to protein synthesis and degradation then follows rate equations of the form 
\begin{eqnarray}
  \frac{d}{dt} A_{\rm tot} & = & \alpha_A P_A(B_2,R_2) - 
  \lambda_p A_{\rm tot} \nonumber \\
  \frac{d}{dt} B_{\rm tot} & = & \alpha_B P_B(A_2,RS) - 
  \lambda_p B_{\rm tot} \;. 
  \label{EQNreduced_model}
\end{eqnarray}
Here, we have assumed that protein degradation occurs at the constant rate $\lambda_p$. In contrast, the synthesis of proteins A and B is regulated. Their maximal synthesis rates are denoted by $\alpha_A$ and $\alpha_B$, while the form of the regulatory control is described by the promoter activity functions $P_A(B_2,R_2)$ and $P_B(A_2,RS)$. The promoter activity is the fraction of time the promoter is not blocked by a repressor and thereby free to bind RNA polymerase ($P_A$ and $P_B$ take on dimensionless values between 0 and 1) \citep{Bintu_CurrOpinGenetDev_05a,Bintu_CurrOpinGenetDev_05b}. The arguments of $P_A$ and $P_B$ are the concentrations of the dimeric repressors which down-regulate the transcription. Within the thermodynamic model for transcription regulation \citep{Bintu_CurrOpinGenetDev_05a,Bintu_CurrOpinGenetDev_05b}, the promoter activity function $P_A$ takes the form 
\begin{equation}
  \label{PAFA}
  P_A(B_2,R_2) = \Big(1+ \frac{B_2}{K_{O_B}}\Big)^{-2} \Big(1+\frac{R_2}{K_{O_{R_2}}}\Big)^{-1} 
\end{equation}
and similarly, 
\begin{equation}
  \label{PAFB}
  P_B(A_2,RS) = \Big(1+ \frac{A_2}{K_{O_A}}\Big)^{-2} \Big(1+\frac{RS}{K_{O_{RS}}}\Big)^{-1} \;.
\end{equation}
Here, the $K$'s denote the equilibrium dissociation constants {\it in vivo} \citep{Gerland_PNAS_02} for the dimer-operator interaction. To achieve the cooperativity required for the bistability of the toggle switch \citep{Cherry_JTheorBiol_00}, we have assumed two binding sites each for the repressor dimers $A_2$ and $B_2$, which is reflected in the square of the first factor \citep{Bintu_CurrOpinGenetDev_05a,Bintu_CurrOpinGenetDev_05b}. Note that the dimer concentrations $A_2$ and $B_2$ in Eqs.~(\ref{PAFA}) and (\ref{PAFB}) must be expressed in terms of the total protein concentrations $A_{\rm tot}$ and $B_{\rm tot}$ to close the rate equations (\ref{EQNreduced_model}), see {\it Supplementary Material} for the explicit expressions. Similarly, the concentrations of the control proteins, $R_{2}$ and $RS$, are functions of the total protein concentrations $R_{\rm tot}$ and $S_{\rm tot}$, which we use to quantify the strengths of the read and input signals\footnote{For simplicity, we do not include the possible homodimerization of S in our model (434 repressor and its mutant $434R[\alpha 3(P22R)]$ can both homodimerize \citep{Hollis_PNAS_88}). Allowing $\Stwo$ dimer formation only reduces the effective concentration of S available for heterodimerization and can be compensated for by increasing the expression rate of S.}.

\begin{figure}[h]
\centerline{\includegraphics[width=12cm]{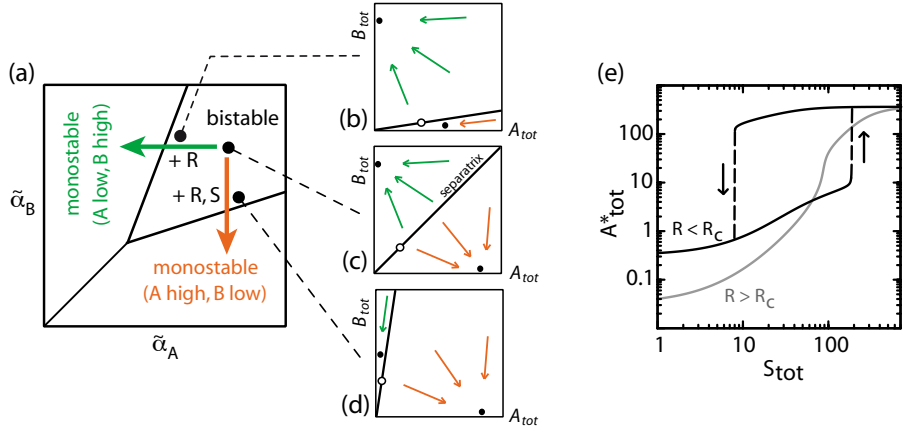}} 
\caption{\label{FIGseparatrix} 
Working Principle of the Conditional Memory Circuit. }
\end{figure}

Without read signal ($R_{tot}=0$), the second factor on the right hand side of (\ref{PAFA}) and (\ref{PAFB}) disappears and the conditional memory circuit behaves like the regular toggle switch \citep{Gardner_Nature_00}. The toggle switch can show three different behaviors depending on the ratio of the maximal promoter activities $\alpha_A$ and $\alpha_B$: The switch is bistable only when $\alpha_A$ and $\alpha_B$ are similar and both are sufficiently strong; otherwise it is monostable, either always ON (A high, B low) or always OFF (A low, B high), see Fig.~\ref{FIGseparatrix}(a). In the conditional memory circuit, the maximal promoter activities effectively get replaced by $\tilde{\alpha}_A = \alpha_A/(1+R_2/K_{O_{R_{2}}})$ and $\tilde{\alpha}_B = \alpha_B/(1+RS/K_{O_{RS}})$. Hence, variation in the concentrations of the control proteins R and S effectively change the maximal promoter activities, and thus can be interpreted as regulated shifts within the state space of the toggle switch. Without read signal, the conditional memory circuit is in the bistable regime and remembers its state. When it receives a read signal, but no S signal, it moves into the monostable low-A regime as indicated by the green arrow in Fig.~\ref{FIGseparatrix} (a). In contrast, when it receives an R and S signal together, it moves into the monostable high-A regime as indicated by the red arrow. The mechanism underlying these operations is illustrated in Figs.~\ref{FIGseparatrix}~(b-d), which represent the dynamic properties of the circuit at the three indicated points in the state diagram Fig.~\ref{FIGseparatrix}~(a):  At any given time, the state of the two-gene circuit is specified by the two concentrations $A_{\rm tot}$ and $B_{\rm tot}$ and its dynamics is therefore represented by trajectories in the A-B-plane. These trajectories tend to the stable fixpoints shown as filled circles in Figs.~\ref{FIGseparatrix}~(b-d). Within the bistable regime, there are two stable fixpoints and the state space is divided by a separatrix into two basins of attraction for these fixpoints (the empty circle indicates the unstable fixpoint). When a read signal is given, this separatrix tilts either towards the A-axis or the B-axis, depending on the signal S. Thereby, the basin of attraction of one of the fixpoints is eliminated, so that the circuit is ultimately forced into the remaining fixpoint. The upward/downward tilting of the separatrix is the physical working principle underlying the two functional operations of Fig.~\ref{FIGscheme}(f). This simple picture holds only within the reduced model (\ref{EQNreduced_model}), but we will see that the qualitative behavior survives in a more realistic quantitative description.

\subsection{Switchable hysteresis}

A key property of the conditional memory circuit is to be sensitive to the input S, if the read signal R is high, and insensitive to S when R is low. With our circuit design, this conditional sensitivity results from a hysteresis effect, which exists at low R, but is switched off when R is raised (the switchable hysteresis is similar to that of magnetic memory, see the Supplementary Fig.~S3 for a comparison). To illustrate this behavior, Fig.~\ref{FIGseparatrix}~(e) shows the steady-state concentration of gene A as a function of $S_{tot}$. At a low R level (black curve), gene A has two steady-state values, one low and one high, over most of the plotted range of $S$ (bistable regime). When $S$ is raised from a low value, gene A remains in the lower steady state and switches to the higher steady state only at the upper end of the bistable regime. Conversely, if $S$ is lowered from maximal expression, A remains in the higher steady state until the lower end of the bistable regime is reached. The grey curve shows the behavior at high R: Over the entire interval, gene A has a unique steady state, and the steady-state concentration increases monotonically with $S$. Hence, the state of gene A follows that of the signal S.

\subsection{Circuit function with realistic parameters and in the presence of noise}

So far we have illustrated the functionality of the conditional memory circuit only schematically. We will now scrutinize the circuit design more closely within a detailed quantitative model, see Fig.~\ref{FIGmodel} and `Materials and Methods«. In particular, we are interested in the timescales of the circuit dynamics given realistic parameters, and in the reliability of the circuit function given the noise in the involved biochemical reactions. For instance, how long does a read signal have to be presented in order to ascertain that the circuit will have stored the signal? As a realistic example we have in mind an implementation of the conditional memory circuit using TetR as protein A, LacR as protein B, the bacteriophage 434 repressor as protein R, and the mutated 434 repressor of Ref.~\citep{Hollis_PNAS_88} as protein S. We use the corresponding experimental parameters when known and typical values in all other cases, see Tables S1 and S2 in {\it Supplementary Material}. 


\subsection{Switching dynamics}

\begin{figure}[h]
\centerline{\includegraphics[width=8cm]{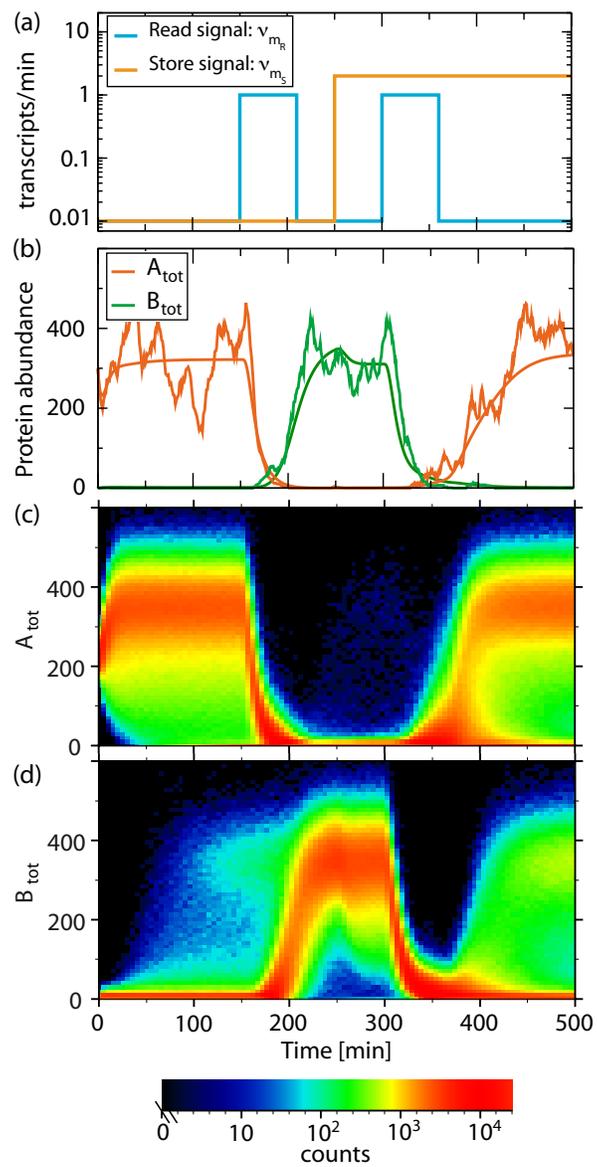}}
\caption{\label{FIGdynamics} 
Response Characteristics of the Conditional Memory Circuit.
}
\end{figure}

To perform its intended function, the conditional memory circuit must respond to time-dependent input signals in the way illustrated in Fig.~\ref{FIGscheme}(f). To verify this response within the quantitative model, we prepare the circuit in the ON state (A high) and then impose time-dependent transcription rates $\transmR(t)$ for R and $\transmS(t)$ for S with shapes as shown in Fig.~\ref{FIGdynamics}(a). The protocol of Fig.~\ref{FIGdynamics}(a) tests the complete set of fundamental circuit operations in the order: (i) remember S high, (ii) read S low, (iii) remember S low, and (iv) read S high. Fig.~\ref{FIGdynamics}(b) shows the response for both genes, A and B, within the deterministic description (smooth green and red curves) as well as exemplary trajectories from a stochastic simulation (rugged green and red curves, respectively). Qualitatively, the curves exhibit the desired behavior: From $t=0-150$~min, when both R and S are in the low state (basal transcription rate of 0.01/min), gene A remains in the ON state. Shortly after the transcription of R is turned on at $t=150$~min, the memory switches to the state of S, i.e. the OFF state. When the transcription of R is stopped at $t=210$~min, gene A remains OFF, even after S is switched to the ON state at $t=250$~min. The change in the state of S affects gene A only after the transcription of R is turned on again at $t=300$~min. 

From Fig.~\ref{FIGdynamics}(b), we can read off the characteristic time required to flip the switch: The duration from the onset of the read signal to the point where gene A and B have reached the same expression level is about 35~min when the switch is flipped from ON to OFF and $\approx 1$~hour in the reverse direction. Since these timescales are relatively long compared to all biochemical timescales in our model, it is natural to ask what mechanism sets them. We found that the flipping timescale increases approximataly linearly with the protein half-life (data not shown). Indeed, the fundamental timescale for all changes in protein concentrations is set by the degradation rate, i.e. the half-live of 5 minutes in the present case. For the switch to flip, the concentration of the dominant repressor has to drop below the apparent threshold for the binding to its operator sites\footnote{We define the {\it apparent} binding threshold as the concentration of transcription factor, that is needed to reduce the promoter activity to 50\% of its maximal value. It is not neccessarily equal to the equilibrium dissociation constant K, but rather depends on the explicit expression of the promoter activity function.}, which amounts to more than a 50-fold change, given the apparent binding thresholds of about 4~nM and 3~nM for the $A_{2}$ and $B_{2}$ operator sites, respectively. Hence, a minimum of 6 protein half-lives is required to flip the switch. The asymmetry between the two switching directions is mainly caused by the asymmetry of the regulatory front end as will become clear below. \\


\subsection{Noise-induced errors} 

The stochastic trajectory in Fig.~\ref{FIGdynamics}(b) displays significant fluctuations, but behaves otherwise similar to the noise-less deterministic curve. While this is the case for most stochastic trajectories, some deviate qualitatively from the intended response. To examine these noise-induced errors, we performed 50000 simulation runs for the same input protocol of Fig.~\ref{FIGdynamics}(a) and determined the probability densities $p(A_{\rm tot},t)$ and $p(B_{\rm tot},t)$ of the total protein concentrations. The time-evolution of these densities is shown in Figs.~\ref{FIGdynamics}(c) and (d), respectively. Noise-induced errors are clearly visible, for instance in the time interval from 0 to 150 min, where a small part of the density of $A_{\rm tot}$ ``leaks'' into the low state, while some of the $B_{\rm tot}$ density leaks into the high state. Since the read signal is absent in this time interval, the leakage corresponds to noise-induced spontaneous flipping of the conditional memory circuit with a concomitant loss of the stored information. The same effect occurs in the standard genetic toggle switch, for which it has been thoroughly characterized \citep{Kepler_BiophysJ_01,Aurell_PRL_02,Allen_PRL_05}. In our case, the average lifetime of the ON state is 32 hours and that of the OFF state is around 600 hours, i.e. both lifetimes are long compared to the timescale for controlled switching. However, the new aspect of the conditional memory circuit is that biochemical noise leads to two additional types of noise-induced errors: During a read pulse (high R), the switch may not flip even though it is triggered to do so (false negative), or the switch may flip, even though it was already in the correct state (false positive).

\begin{figure}[h]
\centerline{\includegraphics[width=15cm]{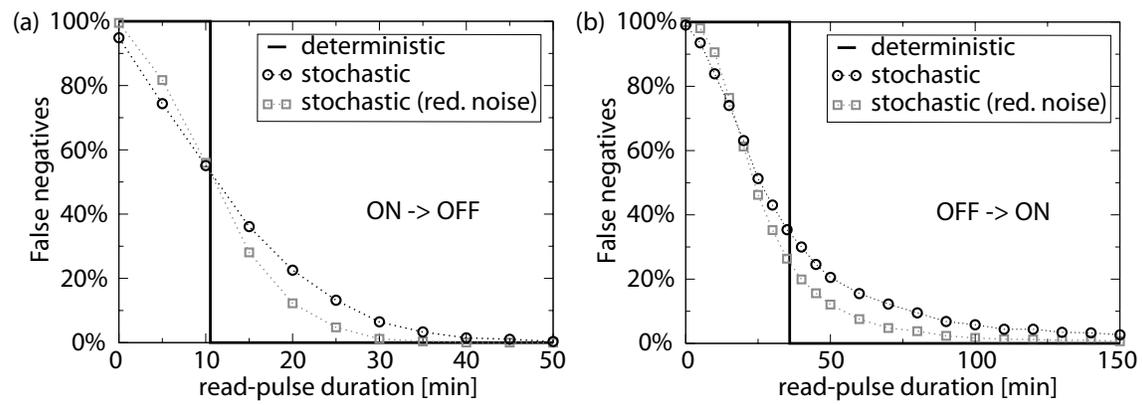}}
\vspace*{0.3cm}
\caption{\label{FIGsuccessratio}
Switching Errors.
}
\end{figure}

False negative errors are visible in Figs.~\ref{FIGdynamics}(c) and (d) at $t\approx 400$~min right after the second read pulse, where a certain fraction of the $A_{\rm tot}$ density remains in the low state, while the same fraction of the $B_{\rm tot}$ density erroneously ends up in the high state. In contrast, for the inverse switching direction (after the first read pulse), we observe hardly any false negatives. To quantify the false negatives, we determine the fraction of false negatives at different read pulse durations for both switching directions\footnote{We allow for a relaxation time of 60~min after the end of the read pulse and then determine the error fraction. Since the rate of spontaneous flipping is very low, the result depends only very weakly on the precise value of the relaxation time, provided it is not too short.} (circles in Fig.~\ref{FIGsuccessratio}). We observe that the error fraction decreases rapidly with increasing read pulse duration. For long read pulses, it drops below 0.5~\% when switching OFF and below 4~\% when switching ON. To probe the quantitative effect of the intrinsic expression noise on the error fraction, we reduced the translation rate tenfold and simultaneously increased the transcription rate by the same factor for all genes (A, B, R, and S). This effectively reduces the burst size, i.e. the average number of proteins produced per transcript \citep{Thattai_PNAS_01}, while keeping the average protein levels and the characteristic timescales invariant. The resulting error fraction (grey squares in Fig.~\ref{FIGsuccessratio}) displays an increased sensitivity to the read pulse duration, so that the error fraction drops more quickly with the duration. If noise effects were completely absent, the circuit would display a sharply defined 'toggle time', i.e. a threshold read pulse duration, above which the circuit is driven into the basin of attraction of the switched state, and below which it remains in the original state. These characteristic toggle times are indicated by the solid lines in Fig.~\ref{FIGsuccessratio} for each switching direction. 

Note that the toggle times for the two switching directions in Fig.~\ref{FIGsuccessratio} (a) and (b) differ by a factor of more than three. This asymmetry cannot be explained by the small difference between the apparent binding thresholds of the $A_{2}$ and $B_{2}$ operator sites (see above). Instead, the primary cause is the intrinsically asymmetric design of the regulatory front end: to turn ON the switch, both S and R proteins are required, which not only form the $RS$ heterodimers, but also a noticeable amount of homodimers $R_2$. These lead to an unwanted partial suppression of gene A, which should be turned ON. Therefore, the protein synthesis of A is reduced, so that it takes a longer time to reach the threshold concentration required to flip the switch\footnote{Note that while the time to reach the steady state concentration only depends on the degradation rate, the time to reach a certain threshold concentration also depends on the synthesis rate.}. In contrast, when the switch is turned OFF, the level of S is low and the R proteins form almost exclusively homodimers, so that there is practically no repression of gene B. 

Compared to the false negative errors, the rate of false positive errors is generally small. To quantify this rate, we prepared the circuit in the ON state, turned the signal S on, and then applied a read pulse to test for false positives in the ON state, and similarly for those in the OFF state, see Fig.~S4 in {\it Supplementary Material}. In the latter case, the fraction of false positives is very small at $<0.1$~\%, while it is more significant in the ON state at $< 4$~\% (see Fig.~S5 in {\it Supplementary Material}), again due to the asymmetry of the regulatory front end. 


\subsection{Adaptation of the circuit to different input signal levels} 

Above, we varied the duration of the read signal, but assumed a given fixed set of concentrations for the high and low levels of the R and S proteins. However, when the conditional memory circuit is embedded into the cellular environment, it must be adjustable to work with a variety of input signals, the level of which will depend on the specific context: in one situation an S concentration of 50 molecules per cell might correspond to the ON state of a signaling process, while in another situation this could be the basal level in the OFF state. For a given set of circuit parameters, there exists a certain threshold concentration (or ``set point'') for S, below which the memory flips to the OFF state and above which it flips to the ON state when a read signal is given. Similarly, there is a set point for the read signal, above which the circuit reads the input and below which it ignores the input. For the circuit design to be versatile, these set points must be {\em programmable}, so that they can be adjusted to lie between the typical high and low levels of S and R, respectively. For our circuit design this can be achieved by exploiting the programmability of operator binding affinities through simple changes in their nucleotide sequence \citep{Gerland_PNAS_02}: the response of the toggle switch of Fig.~\ref{FIGmodel} (top) to the regulatory front end (bottom) critically depends on the binding thresholds of the $\Rtwo$ and RS binding sites. Fig.~S6 in the {\it Supplementary Material} shows that variation of these binding thresholds allows to adjust the set points for S and R over 1-2 orders of magnitude. Thus, we expect that the circuit can easily be adapted to work under a wide range of conditions.

\section{Discussion}

\subsection{Circuit function from a biological perspective} 

In this paper, we described the design of a simple genetic circuit to implement conditional memory in bacteria, and characterized the circuit dynamics theoretically. The additional layer of control (compared to the basic toggle switch) dictates the condition (level of R) by which transient information (level of S) may be stored in the toggle switch. Conditional memory would then enable cells to manipulate information ``collected'' under different conditions at different times. Such capabilities can provide selective advantages to microorganisms in time varying environments \citep{Kussell_Science_05}. 
For instance, under repeated cycles of famine and feast, bacteria which can remember certain environmental traits during feast may formulate better survival strategies at the time of famine. In this functional context, the transcription of S may be controlled by a ``sensor module'' \citep{Kobayashi_PNAS_04} responding e.g. to the light intensity \citep{Levskaya_Nature_05} or density of bacteria \citep{Bassler_Curr_Opin_Microbiol_99}, while the transcription of R may be driven by metabolic or growth regulators that signal the internal state of the cell. More generally, it has been shown on theoretical grounds by \cite{Kussell_Science_05} that cell populations can benefit from stochastic phenotypic switching with memory, where the individuals remember the last few phenotypic switches that occurred in their ancestral history. This benefit arises when the fluctuations of the environment exhibit longer correlations. The implementation of stochastic phenotypic switching with memory would require advanced signal processing capabilities for which the functionality of the conditional memory circuit is fundamental (see also the discussion further below). 

Crucial ``device properties'' of conditional memory include a rapid timescale of active switching, a slow spontaneous loss of the memory content, and a versatile interface to the input signals. According to our design and analysis, we expect the circuit to be able to respond rapidly to variations in input signals, on a time scale as short as 30~min (if active protein degradation is used \citep{Karzai_Nat_Struct_Biol_00}). Our analysis also suggests that a broad parameter regime can be found for which the memory is stable to stochastic fluctuations in gene expression for many cell generations. Furthermore, we showed that the circuit can easily be adjusted to function with a wide range of input signal amplitudes (e.g., the transcription rates for R and S), and therefore could be employed as a functional module in many different contexts. 

It is presently not known whether microorganisms do in fact use conditional memory. Our study indicates that they {\em can} implement conditional memory with readily available components. We hope this work will stimulate the construction and experimental characterization of the conditional memory circuit. We feel that its implementation would be a milestone in synthetic biology, which through its forward engineering approach complements the reverse engineering efforts in microbiology.

\subsection{Sequential logic in gene regulation} 

An important motivation for our study was to explore the design of sequential logic in gene regulation. Our case study of the conditional memory circuit permits some general conclusions, which are best appreciated by recalling the basic hierarchy of digital logic elements \citep{Katz_04}. The minimal elements of digital logic are combinatorial logic gates, which assign an output value to two (or more) inputs according to a fixed rule (e.g. AND, OR, XOR, etc). Most tasks in digital electronics require sequential logic elements, which yield an output level that depends not only on the input levels, but also on a stored level. The most basic sequential logic element is an uncontrolled RS-latch (or flip-flop), a circuit with two states, `0' and `1', which can be affected at any instant of time by two separate inputs, which either set it to state 1 or reset it to state 0. Such uncontrolled (or `asynchronous') elements are rarely used in circuits, as they tend to produce unstable circuit behavior. Thus, sequential logic elements are made sensitive to a control (or `enable') signal, typically a clock, which determines whether or not the element responds to the inputs. However, even with the control, the RS-latch has an undesirable ambiguous input condition when both the set and reset signal is presented. In practice, data latches (`D-latches'), which are free of ambiguous input conditions, are used instead. All of the sequential logic elements in digital electronics are constructed by cascading several combinatorial logic gates and introducing feedback paths. 

In gene regulation, combinatorial logic can be flexibly implemented by {\it cis}-regulatory transcription control \citep{Buchler_PNAS_03}. The genetic toggle switch of \cite{Gardner_Nature_00} and its extensions by \cite{Kobayashi_PNAS_04} implement an uncontrolled RS-latch with several variations in the type of the input signals.  
In contrast, our conditional memory circuit implements a data latch. 
In principle, data latches could be constructed by the same strategy as in digital electronic circuits, i.e. by cascading multiple transcription factor genes under combinatorial transcription control and introducing feedback. However, such designs would require many genes and result in slow operation. In this manuscript, we have studied an alternative strategy: The conditional memory circuit is a very compact design, where the properties of typical protein regulators and transcriptional control mechanisms in bacteria are exploited to implement the functionality of a data latch. The central idea behind this seemingly `natural' strategy is to synergistically combine simple protein-protein and protein-DNA interactions to achieve complex function. We performed our analysis for parameters associated with a specific choice for the R/S pair, the 434 repressor and its mutant $434R[\alpha 3(P22R)]$, because their properties had already been characterized quantitatively \citep{Hollis_PNAS_88}. However, we believe such pairs may be readily generated by synthetic design or natural evolution: As demonstrated by \cite{Dmitrova_Mol_Gen_Genet_98}, not only can DNA binding domains of the transcription factors be altered to enable different binding specificities, the dimerization domain can also be manipulated to enable the desired homo- and hetero-dimerization required by our design for the conditional memory circuit. 

\section{Materials and Methods}

\begin{figure}[h]
\centerline{\includegraphics[width=10cm]{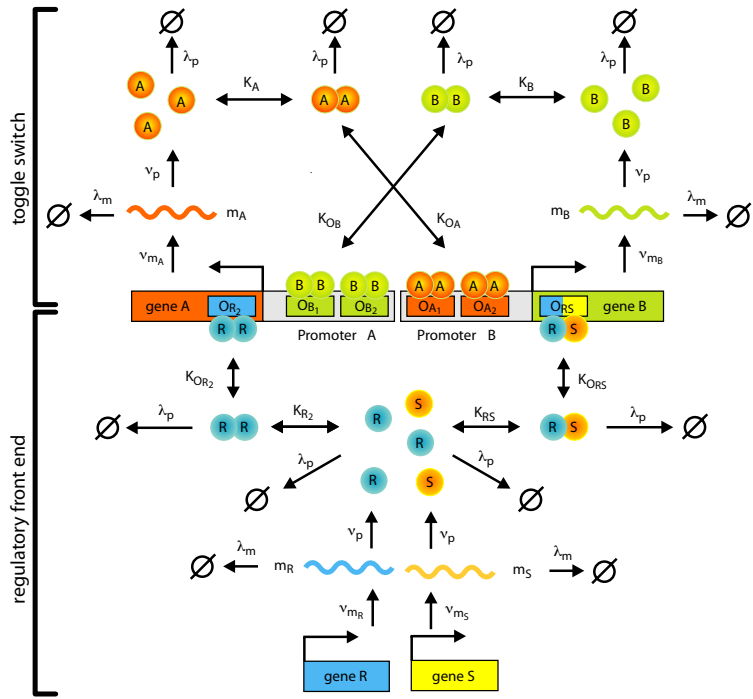}}
\vspace*{0.3cm}
\caption{\label{FIGmodel}
Illustration of the Quantitative Model. 
}
\end{figure}

Our detailed model for the conditional memory circuit describes the explicit dynamics of all biochemical processes shown in Fig.~\ref{FIGmodel}. The top part of Fig.~\ref{FIGmodel} depicts those reactions which are already involved in the toggle switch, whereas the bottom part shows the additional reactions for the conditional memory. Transcription from promoter A and promoter B is regulated by independent\footnote{We assume no direct interaction between the DNA-bound repressors. This assumption is conservative, since cooperative interactions would only help to make the bistability of the circuit more pronounced.} binding of the dimeric repressors $\Atwo$, $\Btwo$, $\Rtwo$, and RS to their respective operator sites. We explicitly describe the dynamics of the repressor-operator interactions to take into account the effect of operator state fluctuations \citep{Kepler_BiophysJ_01}. Transcription occurs only when no repressor is bound to the promoter. In this state, mRNA molecules are produced at the rates $\transmA$ and $\transmB$, respectively (we will assume strong promoters with $\transmA=\transmB=5\,\mathrm{min}^{-1}$). The mRNAs are translated at a rate $\nu_{p}$ and degraded at a rate $\lambda_{m}$ (we fix the average mRNA half-life to a typical value of 3 min \citep{Bernstein_PNAS_02} and assume $\nu_{p}\approx 2.3\,\mathrm{min}^{-1}$ to obtain on average 10 proteins per mRNA molecule, again a typical value \citep{Arkin_Genetics_98,Cai_Nature_06,Yu_Science_06}). The rate for protein turnover, $\lambda_{p}$, is an important parameter, which sets the timescale of the circuit behavior (see below). A rapid response of the memory can be obtained only when degradation is rapid, a constraint that is well known for the toggle switch \citep{Gardner_Nature_00}. We assume that all proteins are actively degraded with half-lives of 5 minutes\footnote{We assume the same degradation rate for protein monomers and dimers, i.e. no cooperative stability \citep{Buchler_PNAS_05}.}, which is usually achieved in synthetic gene circuit experiments by SsrA tags \citep{Elowitz_Nature_00,Basu_Nature_05,Fung_Nature_05}. Our rate constants for the association of monomers and dissociation of dimers are listed in {\it Supplementary Material}, as are the rates for the binding and unbinding of the dimers to their operators\footnote{In our model, dimerization occurs always prior to operator binding. This pathway is consistent with typical parameter values for bacterial transcription factors.}. 

For the control proteins R and S, we describe the expression dynamics in the same way as for A and B, however we consider their transcription rates $\transmR(t)$ and $\transmS(t)$ as time-dependent input signals for the genetic circuit\footnote{Note that we could equally well assume that the transcription of R, S is constant, but their dimerization or their DNA-binding activity is time-dependent, e.g. due to regulation by ligand binding or phosphorylation. However, none of our conclusions are sensitive to the detailed mechanism that in one way or another controls the total concentrations of {\em active} R and S proteins.}. For instance, the protein R could be coupled to the circadian rhythm, with a periodic transcription rate $\transmR(t)$ caused by other regulatory processes in the cell. And the transcription of protein S could be regulated by a signal transduction pathway that is sensitive to a time-dependent external signal. Our aim is to characterize the response of our genetic circuit model to different forms of the input signals $\transmR(t)$ and $\transmS(t)$. To this end, we solve the deterministic rate equations for all reactions by numerical integration, and also perform stochastic simulations of the same reactions using the algorithm of Gillespie \citep{Gillespie_J_Phys_Chem_77}. The dynamics of our detailed model simplifies to that of the reduced model (\ref{EQNreduced_model}) in the limit where (i) all protein concentrations are high, (ii) the dimerization and DNA binding reactions are rapid, and (iii) the mRNA concentrations equilibrate much faster than the protein concentrations. In this limit, the variables $A_{\rm tot}$ and $B_{\rm tot}$ are the only relevant slow degrees of freedom \citep{Bundschuh_BiophysJ_03}. However, in our case these conditions are not met, since the proteins with SsrA tags are degraded almost as rapidly as the mRNA. Hence, we explore the behavior of the full model. 


\section{Acknowledgments}
We thank W. Hillen and G. Koudelka for providing useful information on TetR and 434 repressor. GF is grateful to J. Timmer (University of Freiburg) for guidance and support during his diploma thesis.
NB acknowledges a Burroughs-Wellcome Career Award, TH acknowledges support by the NSF through Grant No. MCB-0417721 and PFC-sponsored Center for Theoretical Biological Physics (Grants No. PHY-0216576 and PHY-0225630), and UG acknowledges an Emmy Noether grant of the {\em Deutsche Forschungsgemeinschaft}. 

\paragraph{Author contributions.} GF performed the simulations. All authors designed the research and wrote the paper. 


\begin{thebibliography}{}

\bibitem[Allen et~al., 2005]{Allen_PRL_05}
Allen, RJ, Warren, PB, and ten Wolde, PR (2005)
\newblock Sampling rare switching events in biochemical networks
\newblock {\em Phys Rev Lett}, 94:018104

\bibitem[Arkin et~al., 1998]{Arkin_Genetics_98}
Arkin, A, Ross, J, and McAdams, HH (1998)
\newblock Stochastic kinetic analysis of developmental pathway bifurcation in
  phage lambda-infected {E}scherichia coli cells
\newblock {\em Genetics}, 149:1633

\bibitem[Arney and Fisher, 2004]{Arney_J_Cell_Sci_04}
Arney, KL and Fisher, AG (2004)
\newblock Epigenetic aspects of differentiation
\newblock {\em J Cell Sci}, 117:4355

\bibitem[Aurell and Sneppen, 2002]{Aurell_PRL_02}
Aurell, E and Sneppen, K (2002)
\newblock Epigenetics as a first exit problem
\newblock {\em Phys Rev Lett}, 88:048101

\bibitem[Bassler, 1999]{Bassler_Curr_Opin_Microbiol_99}
Bassler, BL (1999)
\newblock How bacteria talk to each other: regulation of gene expression by
  quorum sensing
\newblock {\em Curr Opin Microbiol}, 2:582

\bibitem[Basu et~al., 2005]{Basu_Nature_05}
Basu, S, Gerchman, Y, Collins, CH, Arnold, FH, and Weiss, R (2005)
\newblock A synthetic multicellular system for programmed pattern formation
\newblock {\em Nature}, 434:1130

\bibitem[Bernstein et~al., 2002]{Bernstein_PNAS_02}
Bernstein, JA, Khodursky, AB, Lin, P-H, Lin-Chao, S, and Cohen, SN
  (2002)
\newblock Global analysis of {mRNA} decay and abundance in {E}scherichia coli
  at single-gene resolution using two-color fluorescent {DNA} microarrays
\newblock {\em Proc Natl Acad Sci USA}, 99:9697

\bibitem[Bintu et~al., 2005a]{Bintu_CurrOpinGenetDev_05b}
Bintu, L, Buchler, NE, Garcia, HG, Gerland, U, Hwa, T, Kondev, J,
  Kuhlman, T, and Phillips, R (2005a)
\newblock Transcriptional regulation by the numbers: applications
\newblock {\em Curr Opin Genet Dev}, 15:125

\bibitem[Bintu et~al., 2005b]{Bintu_CurrOpinGenetDev_05a}
Bintu, L, Buchler, NE, Garcia, HG, Gerland, U, Hwa, T, Kondev, J, and
  Phillips, R (2005b)
\newblock Transcriptional regulation by the numbers: models
\newblock {\em Curr Opin Genet Dev}, 15:116

\bibitem[Buchler et~al., 2003]{Buchler_PNAS_03}
Buchler, NE, Gerland, U, and Hwa, T (2003)
\newblock On schemes of combinatorial transcription logic
\newblock {\em Proc Natl Acad Sci USA}, 100:5136

\bibitem[Buchler et~al., 2005]{Buchler_PNAS_05}
Buchler, NE, Gerland, U, and Hwa, T (2005)
\newblock Nonlinear protein degradation and the function of genetic circuits
\newblock {\em Proc Natl Acad Sci USA}, 102:9559

\bibitem[Bundschuh et~al., 2003]{Bundschuh_BiophysJ_03}
Bundschuh, R, Hayot, F, and Jayaprakash, C (2003)
\newblock Fluctuations and slow variables in genetic networks
\newblock {\em Biophys J}, 84:1606

\bibitem[Cai et~al., 2006]{Cai_Nature_06}
Cai, L, Friedman, N, and Xie, XS (2006)
\newblock Stochastic protein expression in individual cells at the single
  molecule level
\newblock {\em Nature}, 440:358

\bibitem[Casadesus and D'Ari, 2002]{Casadesus_Bioessays_02}
Casadesus, J and D'Ari, R (2002)
\newblock Memory in bacteria and phage
\newblock {\em Bioessays}, 24(6):512

\bibitem[Cherry and Adler, 2000]{Cherry_JTheorBiol_00}
Cherry, JL and Adler, FR (2000)
\newblock How to make a biological switch
\newblock {\em J Theor Biol}, 203:117

\bibitem[Dmitrova et~al., 1998]{Dmitrova_Mol_Gen_Genet_98}
Dmitrova, M, Younes-Cauet, G, Oertel-Buchheit, P, Porte, D, Schnarr, M,
  and Granger-Schnarr, M (1998)
\newblock A new {LexA}-based genetic system for monitoring and analyzing
  protein heterodimerization in {E}scherichia coli
\newblock {\em Mol Gen Genet}, 257:205

\bibitem[Elowitz and Leibler, 2000]{Elowitz_Nature_00}
Elowitz, MB and Leibler, S (2000)
\newblock A synthetic oscillatory network of transcriptional regulators
\newblock {\em Nature}, 403:335

\bibitem[Fung et~al., 2005]{Fung_Nature_05}
Fung, E, Wong, WW, Suen, JK, Bulter, T, Lee, S, and Liao, JC
  (2005)
\newblock A synthetic gene-metabolic oscillator
\newblock {\em Nature}, 435:118

\bibitem[Gardner et~al., 2000]{Gardner_Nature_00}
Gardner, TS, Cantor, CR, and Collins, JJ (2000)
\newblock Construction of a genetic toggle switch in {E}scherichia coli
\newblock {\em Nature}, 403:339

\bibitem[Gerland et~al., 2002]{Gerland_PNAS_02}
Gerland, U, Moroz, J, and Hwa, T (2002)
\newblock Physical constraints and functional characteristics of transcription
  factor-{DNA} interaction
\newblock {\em Proc Natl Acad Sci USA}, 99:12015

\bibitem[Gillespie, 1977]{Gillespie_J_Phys_Chem_77}
Gillespie, DT (1977)
\newblock Exact stochastic simulation of coupled chemical reactions
\newblock {\em J Phys Chem}, 81:2340

\bibitem[Hollis et~al., 1988]{Hollis_PNAS_88}
Hollis, M, Valenzuela, D, Pioli, D, Wharton, R, and Ptashne, M (1988)
\newblock A repressor heterodimer binds to a chimeric operator
\newblock {\em Proc Natl Acad Sci USA}, 85:5834

\bibitem[Jaenisch and Bird, 2003]{Jaenisch_Nat_Genet_03}
Jaenisch, R and Bird, A (2003)
\newblock Epigenetic regulation of gene expression: how the genome integrates
  intrinsic and environmental signals
\newblock {\em Nat Genet}, 33 Suppl:245

\bibitem[Karzai et~al., 2000]{Karzai_Nat_Struct_Biol_00}
Karzai, AW, Roche, ED, and Sauer, RT (2000)
\newblock The {SsrA-SmpB} system for protein tagging, directed degradation and
  ribosome rescue
\newblock {\em Nat Struct Biol}, 7:449

\bibitem[Katz and Boriello, 2004]{Katz_04}
Katz, R and Boriello, G (2004)
\newblock {\em Contemporary Logic Design}
\newblock Prentice Hall

\bibitem[Kepler and Elston, 2001]{Kepler_BiophysJ_01}
Kepler, TB and Elston, TC (2001)
\newblock Stochasticity in transcriptional regulation: Origins, consequences,
  and mathematical representations
\newblock {\em Biophys J}, 81:3116

\bibitem[Kobayashi et~al., 2004]{Kobayashi_PNAS_04}
Kobayashi, H, Kaern, M, Araki, M, Chung, K, Gardner, TS, Cantor, CR,
  and Collins, JJ (2004)
\newblock Programmable cells: Interfacing natural and engineered gene networks
\newblock {\em Proc Natl Acad Sci USA}, 101:8414

\bibitem[Kussell and Leibler, 2005]{Kussell_Science_05}
Kussell, E and Leibler, S (2005)
\newblock Phenotypic diversity, population growth, and information in
  fluctuating environments
\newblock {\em Science}, 309:2075

\bibitem[Levskaya et~al., 2005]{Levskaya_Nature_05}
Levskaya, A, Chevalier, AA, Tabor, JJ, Simpson, ZB, Lavery, LA,
  Levy, M, Davidson, EA, Scouras, A, Ellington, AD, and Marcotte, EM
  (2005)
\newblock Synthetic biology: engineering {E}scherichia coli to see light
\newblock {\em Nature}, 438:441

\bibitem[Thattai and van~Oudenaarden, 2001]{Thattai_PNAS_01}
Thattai, M and van~Oudenaarden, A (2001) 
Intrinsic noise in gene regulatory networks 
{\em Proc Natl Acad Sci USA}, 98:8614

\bibitem[Trusina et~al., 2005]{Trusina_PLoS_Comput_Biol_05}
Trusina, A, Sneppen, K, Dodd, IB, Shearwin, KE, and Egan, JB (2005)
\newblock Functional alignment of regulatory networks: a study of temperate
  phages
\newblock {\em PLoS Comput Biol}, 1:e74

\bibitem[Yu et~al., 2006]{Yu_Science_06}
Yu, J, Xiao, J, Ren, X, Lao, K, and Xie, XS (2006)
\newblock Probing gene expression in live cells, one protein molecule at a
  time
\newblock {\em Science}, 311:1600

\end{thebibliography}

\section{Captions}

\subsection{Caption to Fig.~\ref{FIGscheme}:} 
{\bf Design and Function of the Toggle Switch Compared to the Conditional Memory Circuit.}\\
In the toggle switch (a) of \cite{Gardner_Nature_00}  the bistable circuit of two mutually repressing genes A, B is controlled by two inducers $I_{1}$, $I_{2}$, effectively implementing the rules (b). The diagrams (c) illustrate how the switch is ``set'' to the ON state, i.e. high (HI) expression of A, and ``reset'' to the OFF state, i.e. low (LO) expression of A, by pulses of $I_{1}$ and $I_{2}$, respectively. In contrast, the conditional memory circuit (d) is regulated by the transcription factors R and S. They form hetero- and homodimers RS and $\Rtwo$ repressing the transcription of gene A and B, respectively. Effectively, the circuit remembers the expression state of S during the last pulse of R. Hence, R functions as a read signal for the information contained in S, as illustrated in the table (e) and the diagrams (f).

\subsection{Caption to Fig.~\ref{FIGseparatrix}:}
{\bf Working Principle of the Conditional Memory Circuit.}\\
When the concentrations of the control proteins R and S change, the circuit moves in the state diagram (a) of the toggle switch, which displays either bistable or monostable behavior depending on the (effective) promoter strengths $\tilde{\alpha}_{A}$, $\tilde{\alpha}_{B}$ of gene A and B. The control proteins R and S tilt the separatrix, which separates the basins of attraction of the two stable fixpoints in the A-B-plane. The orientation of the separatrix is illustrated in (b-d) for three different combinations of R and S. When the circuit reaches the borderline to a monostable regime in (a), one of the stable fixpoints (filled circles) ``annihilates'' with the unstable fixpoint (empty circle). 
(e) The dependence of the steady-state level of gene A on the level of S. At a low R level ($R<R_{c}$, black curve), gene A has two steady-state values over most of the plotted range of S (bistable regime). At a high R level ($R>R_{c}$, grey curve), the bistable regime disappears, and the state of gene A always reflects that of the signal S (the shown curves are obtained with the detailed model as described in `Materials and Methods', and the corresponding contour lines in the state diagram are shown in the Supplementary Fig.~S3).

\subsection{Caption to Fig.~\ref{FIGmodel}:}
{\bf Illustration of the Quantitative Model.} \\
The model comprises the processes of transcription, translation, dimerization, operator binding and degradation of mRNA and proteins. All reactions are shown together with their associated reaction rate or, in the case of reversible reactions like dimerization or protein-DNA binding, their respective equilibrium constant (our results are based on simulations of the full dynamics). \\The upper part depicts the toggle switch module, in which the dimeric proteins of one species can bind to the promoter region of the other one. As soon as one of the two operator sites in either of the promoter regions is occupied, downstream transcription is inhibited. Although we do not consider cooperative interactions of adjacently bound transcription factors, the integration of two binding sites for both A and B is essential for the emergence of bistability.\\
The lower part shows the regulatory front end dictating the state of the toggle switch via two additional binding sites for $\mathrm{R_2}$ and RS downstream of the transcriptional start sites of genes A and B, respectively. The two inputs to the circuit are the transcription rates $\nu_{m_R}$ and $\nu_{m_S}$ of R and S.

\subsection{Caption to Fig.~\ref{FIGdynamics}:}
{\bf Response Characteristics of the Conditional Memory Circuit.}\\
The time-dependent response of the circuit to the input signals (a) is shown in (b-d). The dark red and dark green line in (b) show the total concentrations $A_{\mathrm{tot}}$ and $B_{\mathrm{tot}}$ calculated from the deterministic rate equations, whereas the light red and green lines are obtained from a single stochastic simulation run. The time evolution of the probability densities for $A_{\mathrm{tot}}$ and $B_{\mathrm{tot}}$ are shown in (c) and (d), respectively (the densities are obtained from 50000 stochastic simulation runs; the color codes for the observed number of trajectories inside each bin).

\subsection{Caption to Fig.~\ref{FIGsuccessratio}:}
{\bf Switching Errors.}\\
The fraction of false negative errors as a function of the read pulse duration for (a) switching into the OFF state and (b) into the ON state. The circles indicate data from stochastic simulations with realistic parameters, while the squares show the results for a reduced noise level (obtained with a tenfold reduced translation rate together with a tenfold increased transcription rate, effectively reducing the burst size while keeping the protein levels constant). The solid lines indicate the minimum read pulse duration for switching in the absence of noise (deterministic description).

\newpage
\section{Figures}

\end{document}